\documentclass[final,5p,times,twocolumn]{elsarticle}

\newcommand{\ba}{\begin{eqnarray}}
\newcommand{\ea}{\end{eqnarray}}
\newcommand{\be}{\begin{equation}}
\newcommand{\ee}{\end{equation}}
\newcommand{\eps}{\epsilon}
\pdfoutput = 1

\usepackage{amssymb}

\usepackage{lineno}

\journal{Physics Letters A}
\usepackage{color}
\begin{document}

\begin{frontmatter}



\title{Nonlinear fast growth of water waves under wind forcing}


\author[label1]{Maura Brunetti}
\author[label2]{Nad\`ege Marchiando}
\author[label2]{Nicolas Berti}
\author[label2]{J\'er\^ome Kasparian}

\address[label1]{Institute for Environmental Sciences,
University of Geneva, Route de Drize 7, 1227 Carouge, 
Switzerland}
\address[label2]{GAP-Biophotonics, University of Geneva, 
Chemin de Pinchat 22, 1227 Carouge, Switzerland
}

\begin{abstract}
In the wind-driven wave regime, the Miles mechanism gives an estimate of the growth rate of the waves under the effect of wind. We consider the case where this growth rate, normalised with respect to the frequency of the carrier wave, is of the order of the wave steepness. Using the method of multiple scales, we calculate the terms which appear in the nonlinear Schr\"odinger (NLS) equation in this regime of fast-growing waves. We define a coordinate transformation which maps the forced NLS equation into the standard NLS with constant coefficients, that has a number of  known analytical soliton solutions. Among these solutions, the Peregrine and the Akhmediev solitons show an enhancement of both their lifetime and maximum amplitude which is in qualitative agreement with the results of tank experiments and numerical simulations of dispersive focusing under the action of wind. 

\end{abstract}


\end{frontmatter}


\section{Introduction}
\label{intro}

The investigation of the physical mechanisms for the generation of ocean waves by wind has a long history which starts at the beginning of the 20th century~\cite{1925RSPSA.107..189J} 
and is still ongoing. The problem is highly 
nonlinear~\cite{JanssenBook} and the feedback at the air-water interface between wind and water waves is difficult to study experimentally and theoretically because of turbulence in both fluids. 

The problem can be simplified at first by neglecting currents in the water and by considering the so-called wind-driven wave regime which is characterised by growing seas with wave ages $c_p/u^* < 30$, where $c_p$ is the phase velocity of the water waves and $u^*$ is the friction velocity of wind over water waves~\cite{Sullivan2010}. Direct field measurements of the pressure induced by airflow on waves are rare, thus there is no agreement in the scientific community on the underlying mechanisms leading to wave amplification~(for a review see \cite[Chap.~3]{JanssenBook} and \cite{Sullivan2010}).  
 
In the shear flow model introduced by Miles~\cite{Miles1957,Janssen1991} the rate of energy transfer from the wind to a wave propagating at phase velocity $c_p$ is 
proportional to the wind profile curvature $U''(z_c)$ at the critical height $z_c$ 
where the wind speed 
equals the phase velocity of the wave, $U(z_c) = c_p$. The Miles mechanism 
has been recently confirmed in field experiments, in particular for long waves~\cite{Hristov2003}. 
For a logarithmic velocity profile in the boundary layer, 
the Miles growth rate $\Gamma_M$ results in~\cite{Miles1957,Banner2002,Kharif2010}
\be 
\frac{\Gamma_M}{\omega} = \frac{\Gamma_M}{2\pi f}\equiv \frac{1}{\omega E}\frac{dE}{dt} = \delta\, \alpha  
\left(\frac{u^*}{c_p}\right)^2
\label{Miles}
\ee
where $E$ is the wave energy, $f$ is the frequency of the carrier wave, $\delta = \rho_a/\rho_w$ is 
the density ratio ($1.29\times 10^{-3}$ between air and water), and $\alpha$ is an empirical constant of the order of 32.5 in the wind-driven wave 
regime~\cite{Banner2002}. The pressure 
$P$ induced at the water surface then depends on the surface elevation $\eta$  as follows~\cite{Miles1957,Kharif2010}
\be 
\frac{1}{\rho_w} P(x,t)  = \frac{\Gamma_M}{f} \frac{c_p^2}{2\pi}\,  \eta_x(x,t) 
\label{pressure}
\ee
Typical values of $\Gamma_M/f$ are shown in Fig.~1 of~\cite{Banner2002} (in that figure $\Gamma_M = \gamma$) or in Fig.~1 of~\cite{Farrell2008} (where $\Gamma_M = \beta$) as a function of the wave age $c_p/u^*$. 
They range from $10^{-3}$-$10^{-2}$ for fast-moving waves ($c_p/u^*> 5$) to $10^{-2}$-1 for slow-moving waves and laboratory tank experiments ($c_p/u^*\le 5$). 
Thus, the growth rate can be regarded as a small parameter in the wind-driven wave regime and 
generally it is assumed that  $\Gamma_M/f = O(\epsilon^2)$, where $\eps = a k$ is the wave steepness, $a$ being the amplitude of the vertical water displacement $\eta$ and 
$k$ the wavenumber of the water wave.  For weak-nonlinear waves the steepness is indeed small 
and in ocean waves it is smaller than 0.55, the value for which wave-breaking 
occurs~\cite{Toffoli2010}. 
The case $\Gamma_M/f = O(\epsilon^2)$ gives rise to the following damped/forced 
nonlinear Schr\"odinger equation~\cite{Leblanc2007,Kharif2010,OnoratoProment2012}
\be
i\frac{\partial a}{\partial t} - \frac{1}{8} \frac{\omega}{k^2}\frac{\partial^2 a}{\partial x^2} - 
\frac{1}{2} \omega k^2 |a|^2 a = i \left(\frac{\Gamma_M}{2} - 2\nu k^2\right) a
\label{dampedForced}
\ee 
where  $\nu$ is the kinematic viscosity.
Thus, the case $\Gamma_M/f = O(\epsilon^2)$ describes the quasi-equilibrium 
between wind and damping effects due to viscosity.

In this Letter, we consider the case $\Gamma_M/f= O(\epsilon)$, corresponding to stronger winds, the effect of which overcomes the dissipation due to viscosity. This case turns out to be relevant 
for explaining experimental results obtained in the context of dispersive focusing of waves under the action of wind~\cite{2006EJMF...25..662T,2008NPGeo..15.1023T}.
We will insert the aerodynamic pressure term, given in eq.~(\ref{pressure}),
into the Bernoulli equation evaluated at the ocean surface and we will use the method of multiple scales  to obtain the 
corresponding nonlinear Schr\"odinger equation in the case of fast-growing waves. Due to the 
universality of the NLS equation in many other fields of physics, the considered case can in principle be of interest in other physical 
 situations where the multiple-scale method can be applied and the forcing term is introduced at first order in the development parameter. 
 

\section{Governing equations and the method of multiple scales (MMS)}

We recall here the equations governing the propagation of surface gravity waves in the presence of wind and the main assumptions used in the method of multiple scales for deriving the NLS equation. 
    
At low viscosity the water-wave problem can be set within the framework of potential flow 
theory~\cite{2008Dias} and the two-dimensional flow  of a 
viscous, incompressible fluid is governed by the Laplace equation 
\be 
\nabla^2 \phi = 0
\ee
where $\phi(x,z)$ is the velocity potential.
This equation is solved together with the kinematic boundary condition at the free surface $\eta(x,t)$ 
\be
\eta_t +\phi_x \eta_x - \phi_z = 2\nu \eta_{xx} \qquad{\rm{at}}~~  z=\eta(x,t)
\label{kinem}
\ee
and at the bottom 
\be
\phi_z = 0 \qquad{\rm{at}}~~ z=-H 
\ee
The other boundary condition is given by the Bernoulli equation which at the free surface takes the form 
\be
\phi_t + \frac{1}{2}(\phi_x^2 + \phi_z^2 )+ g\eta =  - \frac{P} {\rho_w}  - 2 \nu \phi_{zz} \qquad{\rm{at}}~~ z=\eta(x,t) 
\label{bernoulli}
\ee
where $g$ is the gravity acceleration and $P$ is the excess pressure at the ocean surface in the presence of wind, given by eqs.~(\ref{pressure})  in the context of the Miles mechanism. 

We use the method of multiple scales (MMS)  to find the terms in the NLS equations which are related to the wind forcing with a growth rate of first order in the wave steepness, 
$\Gamma_M/f= O(\epsilon)$. This method is based on the fact that temporal 
and spatial scales of the 
carrier wave $(1/\omega, 1/k)$ are much smaller than those of the envelope. 
MMS has been used for 
deriving the NLS equation under the assumption of small nonlinearity, $\epsilon = a k \ll1$, and 
narrow spectral width $\Delta k/k  \ll 1$~\cite{1972Hasimoto} and successfully applied for including high-order nonlinear terms~\cite{Slunyaev2005} and constant vorticity in water waves~\cite{2012PhFl...24l7102T}, or in other physical contexts. For example, in the context of  the propagation of optical waves in nonlinear materials~\cite{1994PhRvA..49..574K}, this method is also known as the slowly varying envelope approximation (SVEA)~\cite{2007PhR...441...47C,2007RPPh...70.1633B}.  

The velocity potential $\phi$ and the surface elevation
$\eta$ have the following representations~\cite{Slunyaev2005,2012PhFl...24l7102T}
\ba
\phi & =&  \sum_{j=1}^{\infty} \epsilon^j \phi_j\, , \qquad 
\phi_j = \phi_{j0} + \sum_{n=1}^{j} \phi_{jn}\, {\cal E}^n + c.c. 
\label{mms1} \\
\eta & =&   \sum_{j=1}^{\infty} \epsilon^j \eta_j\, , \qquad 
\eta_j = \eta_{j0} +\sum_{n=1}^{j} \eta_{jn}\, {\cal E}^n + c.c. 
\label{mms2}
\ea
where the second index in the amplitudes $\phi_{jn}, \eta_{jn}$ refers to the harmonics  
\be
{\cal E}^n = \frac{1}{2} \exp[in(kx_0-\omega t_0)]
\ee
The velocity potential at the free surface,  $\phi(x,z=\eta,t)$, is written as a Taylor expansion around $z=0$:
\ba
\phi(x,\eta,t) &=& \sum_{j=0}^{+\infty} \frac{\eta_j}{j!} \partial^j_z \phi\arrowvert_{z=0} \label{Taylor}
\ea
The operators for the derivatives are replaced by sums of operators
 \ba
 \frac{\partial}{\partial t} &=& \frac{\partial}{\partial t_0} + \eps \frac{\partial}{\partial t_1} + \eps^2 \frac{\partial}{\partial t_2} + \ldots
 \ea
 corresponding to fast and slow temporal derivatives,
and analogously for $\partial/\partial x$. 
We use the same notation as in Ref.~\cite{2012PhFl...24l7102T} (note however that the order of indices in the amplitudes $\phi_{jn}, \eta_{jn}$ is inverted). 

\section{Wind-forced NLS equation} 
\label{section:windNLS}

In this section we apply the method of multiple scales for developing  the governing equations
in terms of the expansion parameter $\eps$.   
Terms of linear order in $\eps$ give the dispersion relation $\omega = \sqrt{g\sigma k}$, where 
$\sigma = \tanh(kH)$, and they are not affected by wind forcing. The wind forcing terms appear in the expansion at second order in the following relations:
\ba
\label{eq:cg}
&&\frac{\partial A}{\partial t_1}  + c_g \frac{\partial A}{\partial x_1}  =     \Gamma_M\frac{\sigma A}{2} \\
&&\eta_{21}
= \frac{1}{g} \left (  i \omega D + c_g \frac{\partial A}{\partial x_1}  +   \Gamma_M\frac{ \sigma A}{2}\right) 
\label{eq:eta21new}
\ea
where $A= \phi_{11}|_{z=0}$ and $D=\phi_{21}|_{z=0}$.

At third order, the new terms are $\partial \eta_{21}/\partial t_1$ in the kinetic boundary 
condition (\ref{kinem}), which must be evaluated using eq.~(\ref{eq:eta21new}), and $(\Gamma_M/\omega) c_p^2 (\partial \eta_{11}/\partial x_1 +  \partial \eta_{21}/\partial x_0)$ in the Bernoulli equation at $z=0$, eq.~(\ref{bernoulli}). Including these terms finally 
gives the wind-forced NLS equation in the limit of deep-water waves, $kH\gg1$:
\be
i \frac{\partial a}{\partial t_2} -\beta_1  \frac{\partial^2 a}{\partial x_1^2} - \beta_2
\frac{\partial a}{\partial x_1}-\beta_3 a -M a |a|^2 = -2i\nu k^2 a
\label{NLSwind} 
\ee
where $a = 2iA /c_p$, $\beta_1 = -(dc_g/dk)/2 = \omega/(8k^2)$, 
$\beta_2=  \Gamma_M ( \omega + c_g k)/(2\omega k)= 3\Gamma_M/(4k)$, 
$\beta_3 =  \Gamma_M^2 / (8\omega)$ and $M = \omega k^2/2$.

Note that the equation that we obtain for  $\Gamma_M/f = O(\eps)$ differs, as it should, from the usual equation obtained assuming $\Gamma_M/f = O(\eps^2)$. When 
$\Gamma_M/f = O(\eps^2)$, eq.~(\ref{NLSwind}) reduces to eq.~(\ref{dampedForced}).  
Indeed,  the terms proportional 
to $\Gamma_M$ in eqs.~(\ref{eq:cg})-(\ref{eq:eta21new}) become of higher order.  
Moreover, in the Bernoulli equation at $z=0$, the term $(\Gamma_M/\omega)\, c_p^2 (\partial \eta_{11}/\partial x_1 +  \partial \eta_{21}/\partial x_0)$ becomes  
$(\Gamma_M/\omega)\, c_p^2\, \partial \eta_{11}/\partial x_0$. 
This term in  the forced NLS 
equation reduces for $\Gamma_M/f = O(\eps^2)$ to the term  
$- i\Gamma_M a /2$, which corresponds to the one on the right-hand side of eq.~(\ref{dampedForced}).   As we will see in the next section, 
the two terms in the NLS 
equation~(\ref{NLSwind}) due to wind forcing correspond to a variation of the dispersion term and of the phase of the wave field.

It is interesting to calculate the energy evolution. The Miles growth rate is recovered from the relations at second-order expansion. Indeed, multiplying eq.~(\ref{eq:cg}) by the complex conjugate $a^*$, adding  the obtained 
equation to its complex conjugate and integrating by parts yields 
\be
\frac{dE}{dt_1} = \Gamma_M E
\label{windFirstOrder}
\ee
where the wave energy is defined as 
$E = \int |a|^2 dx_1$ and $\rho = |a|^2$ is the energy density. From the same procedure, but starting from the wind-forced NLS equation~(\ref{NLSwind}) at third-order expansion 
we obtain  
\be
\frac{dE}{dt_2} +\int \frac{\partial j}{\partial x_1}dx_1 = \frac{\beta_2}{\beta_1} \int  j\, dx_1
-4\nu k^2 E
\label{en2order}
\ee
where $j$ is the energy flux 
\be
j = -i \beta_1 \left( \frac{\partial a^*}{\partial x_1} a -  \frac{\partial a}{\partial x_1} a^* \right) 
\ee
The second term in eq.~(\ref{en2order}) disappears if we assume that there are no incoming or outgoing waves at infinity and we get
\be
\frac{dE}{dt_2}  = \frac{3 \Gamma_M}{c_g} \int j\, dx_1  -4\nu k^2 E
\label{windSecondOrder}
\ee
At this third order, the wave energy is dissipated  due to viscosity effects (second term on the right-hand side) and it is amplified (for $\Gamma_M>0$) under the action of wind (first term of the right-hand side).  
For comparison, in the case $\Gamma_M/f = O(\eps^2)$ one obtains 
\be
\frac{dE}{dt_1} = 0, \qquad \frac{dE}{dt_2} = (\Gamma_M - 4\nu k^2 ) E
\ee

\section{Reduction of the wind-forced NLS to the standard NLS}
\label{forcedNLS}

In the previous section we have derived the NLS equation in the case where the growth rate of the wave energy  due to the wind effect is of first order in the wave steepness, while viscosity is of second order, eq.~(\ref{NLSwind}). In this section we define 
a coordinate transformation  to obtain the 
standard NLS equation with constant coefficients in order to use its well-known solutions. 

We neglect the viscosity term (since it was already discussed in~\cite{OnoratoProment2012}) and obtain the following equation 
\be
i \frac{\partial a}{\partial t}  - \beta_1 \frac{\partial^2  a}{\partial x^2}  
- \beta_2 \frac{\partial a}{\partial x} - \beta_3 a - M  a |a|^2 = 0
\label{eq:CorrOrder1}
\ee
where $\beta_1 = \omega/(8k^2)$, 
$\beta_2= 3\Gamma_M/(4k)$, $\beta_3 =  \Gamma_M^2 / (8\omega)$ and $M = \omega k^2/2$.
We scale the envelope amplitude as 
\be
a(x,t) = B(x,t) e^{-i\beta_3 t}
\label{scaleBeta3}
\ee
by changing its phase. 
Thus eq.~(\ref{eq:CorrOrder1}) becomes
\be
i \frac{\partial B}{\partial t}  - \beta_1 \frac{\partial^2  B}{\partial x^2}  
- \beta_2 \frac{\partial B}{\partial x} - M  B |B|^2 = 0
\label{eq:scaled}
\ee

The coordinate transformation which directly reduces eq.~(\ref{eq:scaled}) to the standard NLS equation with constant coefficients is given by 
\be
y = -i\beta_2 t + x, \qquad \tau = t
\label{coordTrans}
\ee
Indeed, after this transformation eq.~(\ref{eq:scaled}) results in
\be
i \frac{\partial B}{\partial \tau}  - \beta_1 \frac{\partial^2 B}{\partial y^2}  -M B |B|^2 = 0
\label{standardNLS}
\ee
Thus we have formally mapped the wind-forced NLS equation, eq.~(\ref{eq:CorrOrder1}), 
into the standard NLS equation, which has a number of known analytical soliton solutions (Peregrine, Akhmediev and Kuznetsov-Ma solutions), which we will discuss in the next section.

Alternatively, 
another coordinate transformation is useful for understanding the physical content of eq.~(\ref{eq:CorrOrder1}). We consider the coordinate transformation $\xi = - e^{-b x}/b$, where 
$b = \beta_2/\beta_1 = 3\Gamma_M/c_g$. The derivatives become
\ba
\frac{\partial B}{\partial x} &=& e^{-b x} \frac{\partial B}{\partial \xi} \\
\frac{\partial^2 B}{\partial x^2} &=& -b e^{-bx } \frac{\partial B}{\partial \xi} + \frac{\partial^2 B}{\partial \xi^2} e^{-2bx}
\ea
and eq.~(\ref{eq:scaled}) reduces to
\be
i \frac{\partial B}{\partial t}  - \beta_1 (\xi b)^2 \frac{\partial^2 B}{\partial \xi^2}  -M B |B|^2 = 0
\label{dispTerm}
\ee
 The factor $(\xi b)^2$  modulates the dispersion term and consequently affects the  focusing properties of the system. 
We check that for $b\to0$, the term $\xi b \to -1$ and  we recover the standard NLS equation.

\section{Soliton solutions}
\label{rw}

Eq.~(\ref{standardNLS}) being the standard NLS equation, its solutions include 
the Peregrine, the Akhmediev, and the Kuzbetsov-Ma solutions. Here we discuss how the 
coordinate transformation (\ref{coordTrans}) modifies these solutions. 
 
We start from the solutions $B(y,\tau)$ of eq.~(\ref{standardNLS}) and we perform the transformation $y = -i\beta_2 t+x$, 
$\tau = t$, where $\beta_2$ is the coefficient of the wind-forcing term, 
$\beta_2  =   3 \Gamma_M /(4k)$. 
Finally, we scale the fields as in eq.~(\ref{scaleBeta3}) to obtain 
$a(x,t) = B(x,t) e^{-i\beta_3 t}$. 

The analytical form of the Peregrine solution~\cite{peregrine,2013PhR...528...47O} therefore becomes 
in the case of fast-growing waves
\ba
&&a(x,t) = a_0 \exp[-i M a_0^2 t-i\beta_3 t]\times \nonumber \\ 
&& \left(\frac{4\beta_1(1-2iMa_0^2 t)}{\beta_1+\beta_1(2Ma_0^2\,t)^2 + 2Ma_0^2\, (x-i\beta_2t)^2} -1\right)
\label{Peregrine}
\ea
In Fig.~1 we show the Peregrine solution for $a_0=1$ and $\eps = 0.1$. In panels $(a)$-$(b)$ 
we show the unperturbed Peregrine solution,~eq.~(\ref{Peregrine}) with $\beta_2 = 0$ and $\beta_3 = 0$, while in 
panels $(c)$-$(d)$ and $(e)$-$(f)$ we show the wind-forced solution with 
$\Gamma_M/f = \eps$ and $\Gamma_M/f = 2\eps$, respectively.
The effect of the wind is to break the symmetry along the direction of wave propagation and to increase the temporal duration of the rogue wave event.  The lifetime and the maximum amplitude of the Peregrine soliton under the effect of wind are shown in Fig.~2 (blue solid and dotted lines). The lifetime of the soliton, defined as the period of time where the rogue wave criterium $a/a_0 > 2.2$ is satisfied, increases as the growth rate increases, while its  maximum amplitude remains constant for 
$\Gamma_M/f < 1.72 \eps$ and slightly increases for $\Gamma_M/f \ge 1.72 \eps$. At
$\Gamma_M/f = 1.72 \eps$ the maximum splits into two peaks symmetrical 
with respect to the $t=0$ line\footnote{It is interesting to note that the values of both amplitudes and lifetimes scale with $\eps$ for the Peregrine soliton, so that the results shown in Figs.~1 and 2 are valid for all $\eps$. This is not true for the Akhmediev soliton.}. Enhancement of both amplitudes and lifetimes of rogue waves under the effect of wind has indeed been observed in tank experiments and in numerical simulations of dispersive focusing~\cite{2006EJMF...25..662T,2008NPGeo..15.1023T} and nonlinear focusing~\cite{TouboulKharif2006}.  In these papers, the parameters are chosen so that the condition 
$\Gamma_M/f = O(\eps)$ is satisfied, ensuring that their results can be compared to the ones presented here. Note that in the case $\Gamma_M/f=O(\eps^2)$ only the soliton maximum amplitude increases, while its lifetime is not affected by the wind~\cite{OnoratoProment2012}. Thus, growth rates of the same order as the steepness are required to reproduce the experimental and numerical results shown in~\cite{2006EJMF...25..662T,2008NPGeo..15.1023T}.

\begin{figure}
\centering
\includegraphics[width=0.23\textwidth]{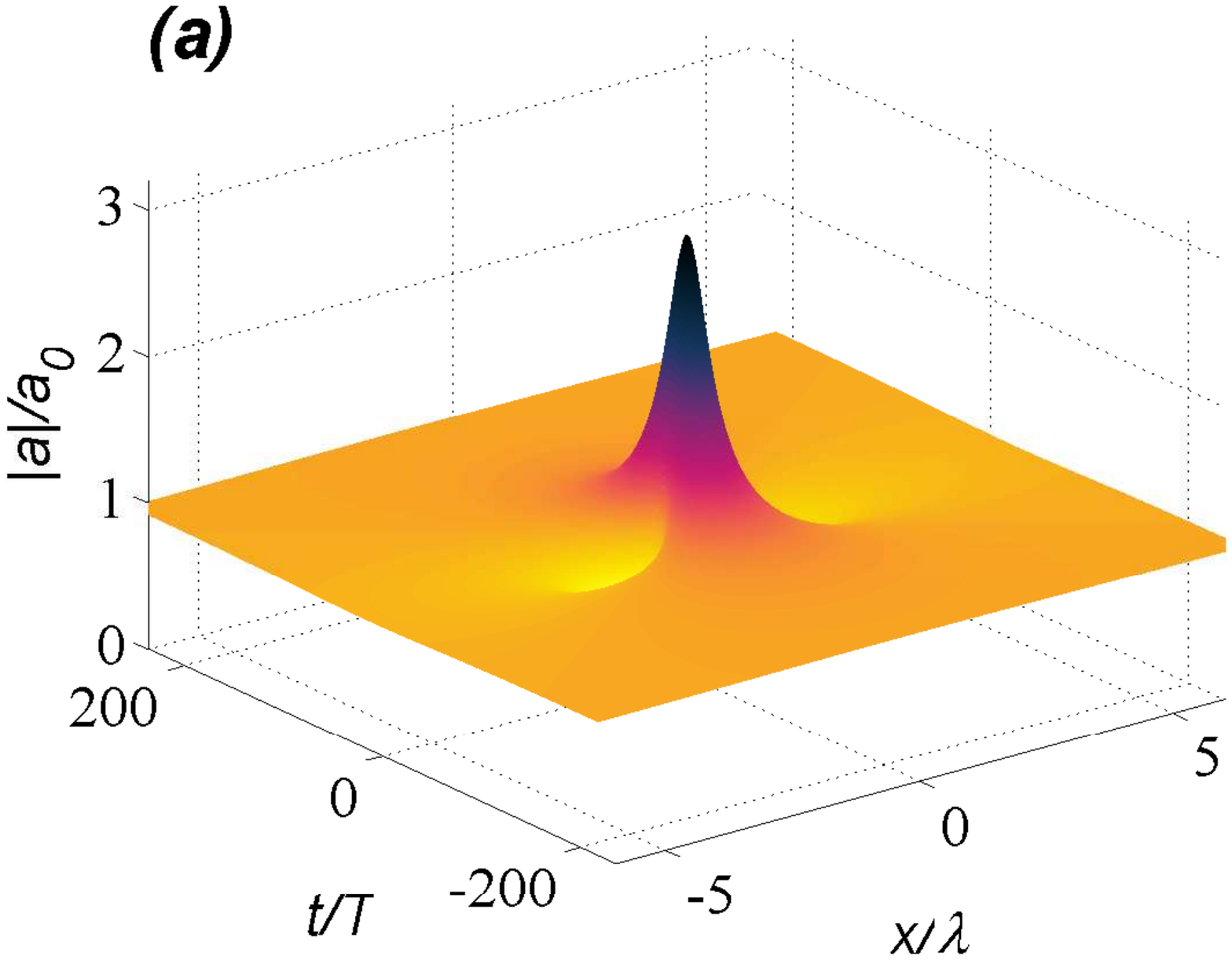}
\includegraphics[width=0.23\textwidth]{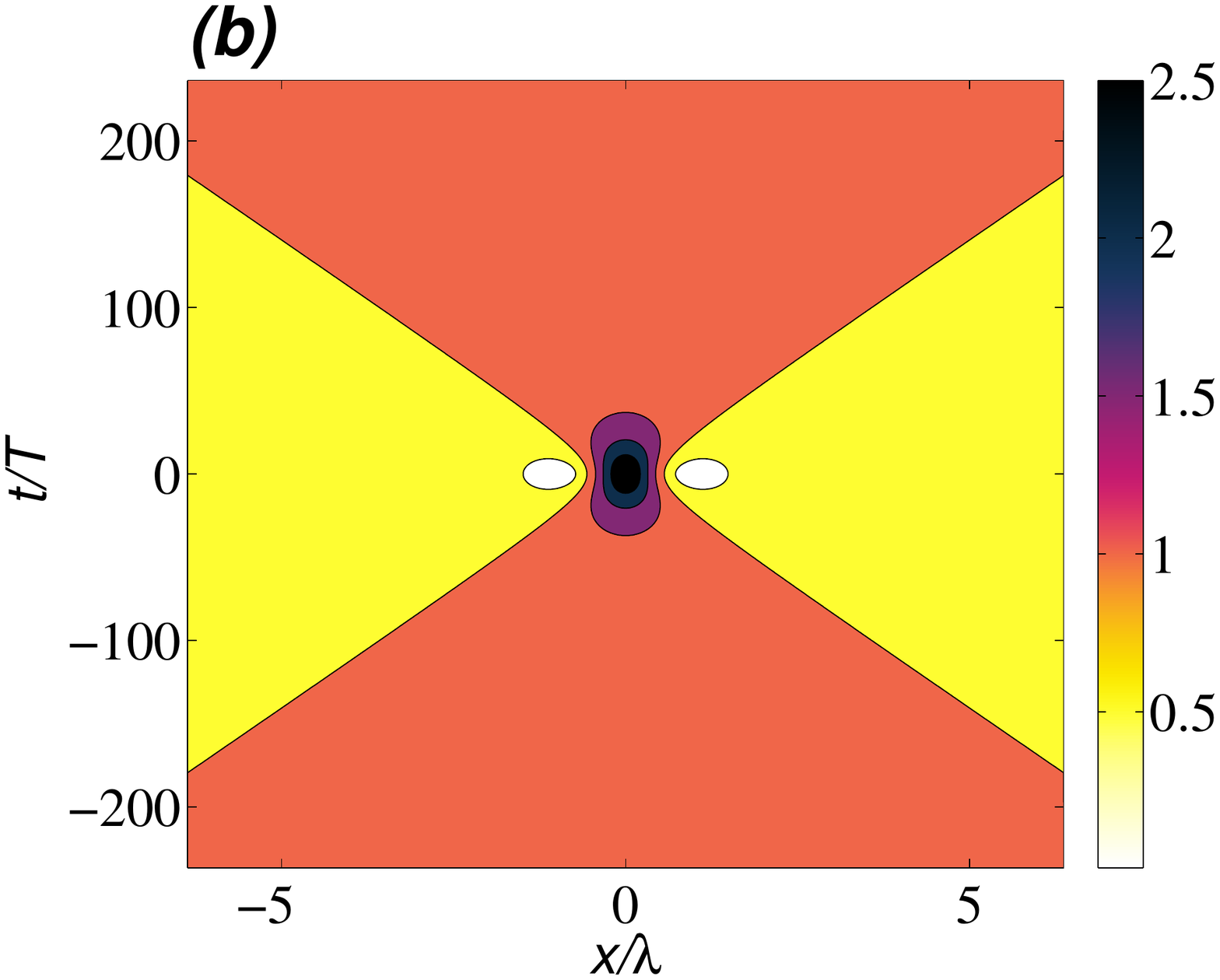}
\includegraphics[width=0.23\textwidth]{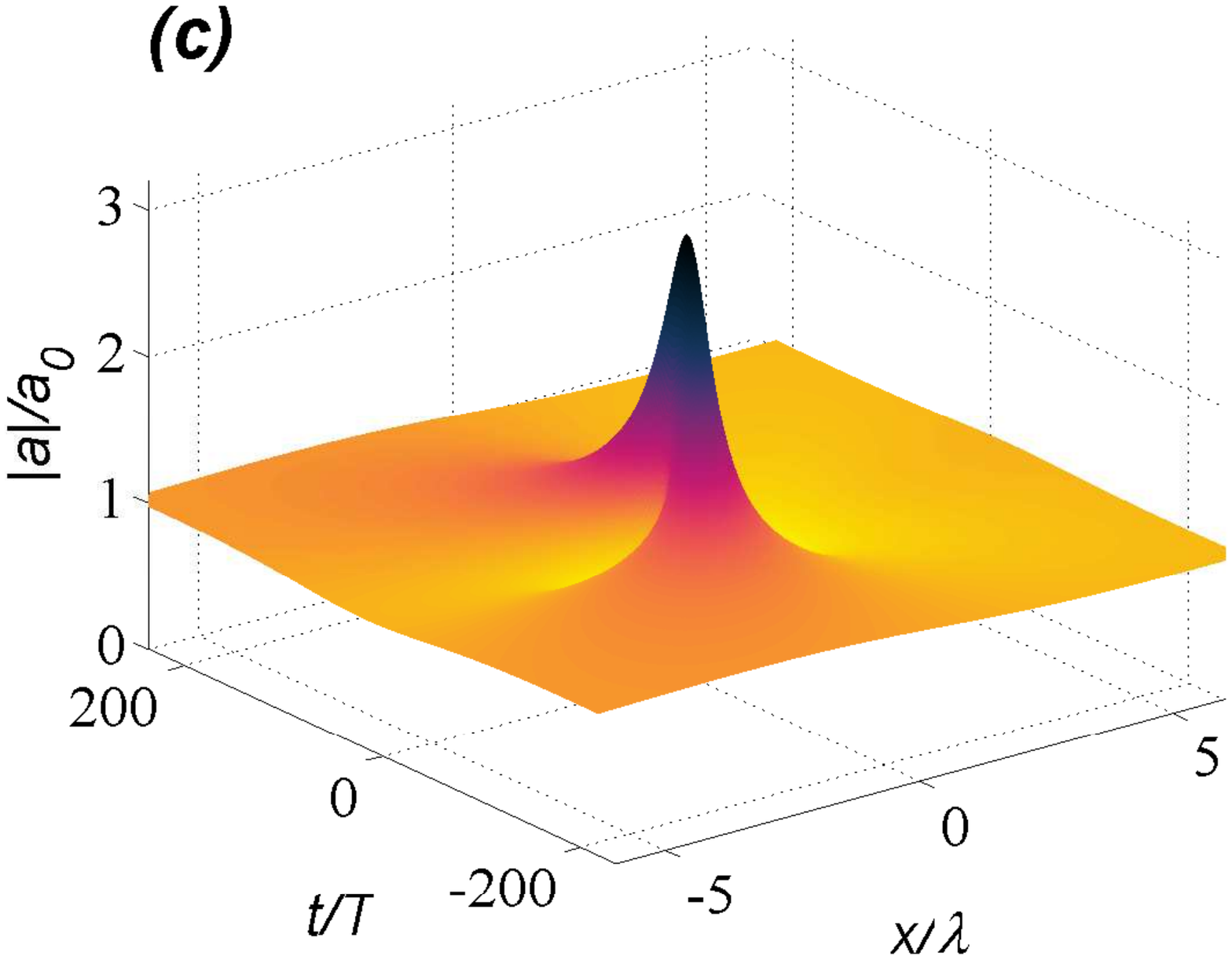}
\includegraphics[width=0.23\textwidth]{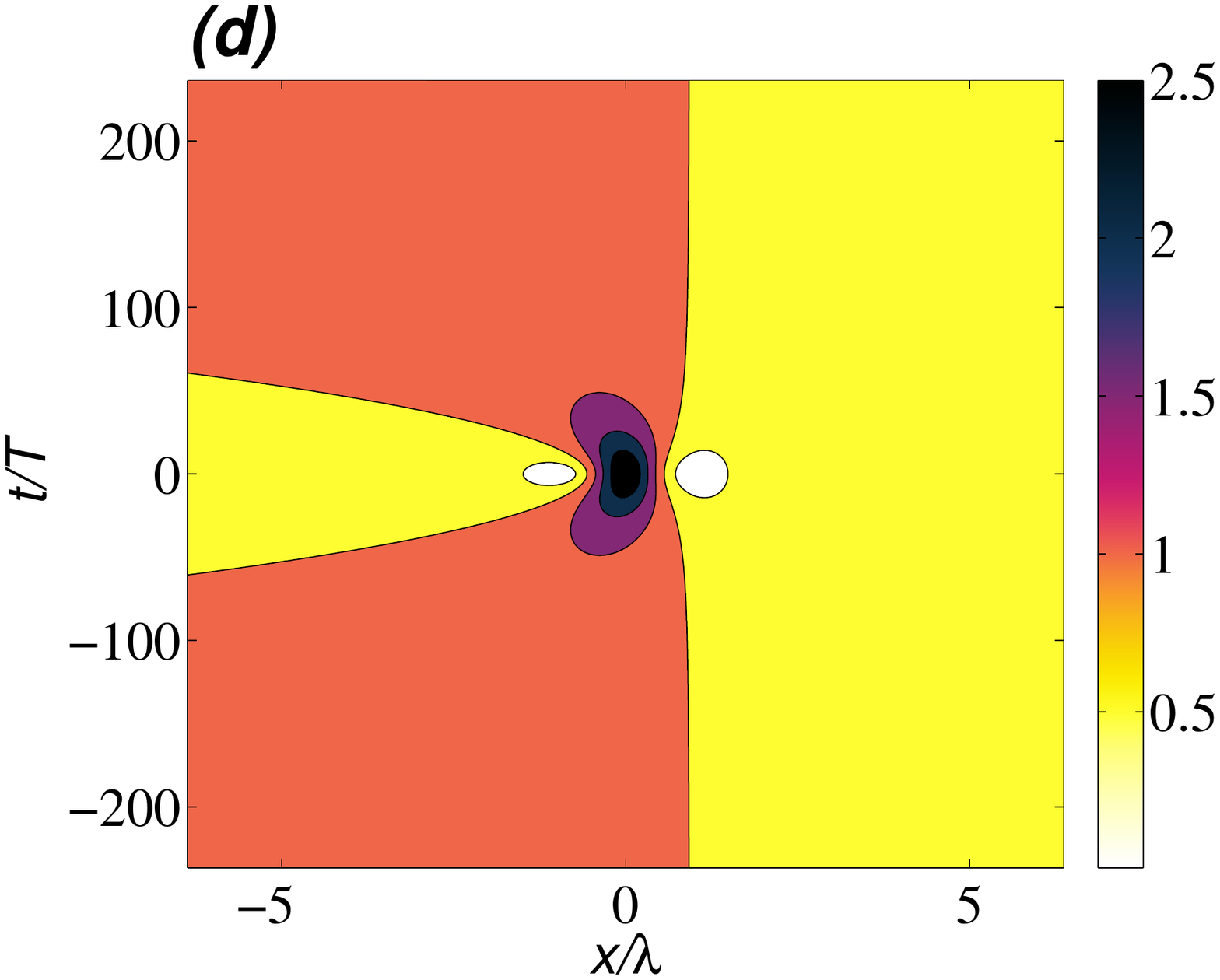}
\includegraphics[width=0.23\textwidth]{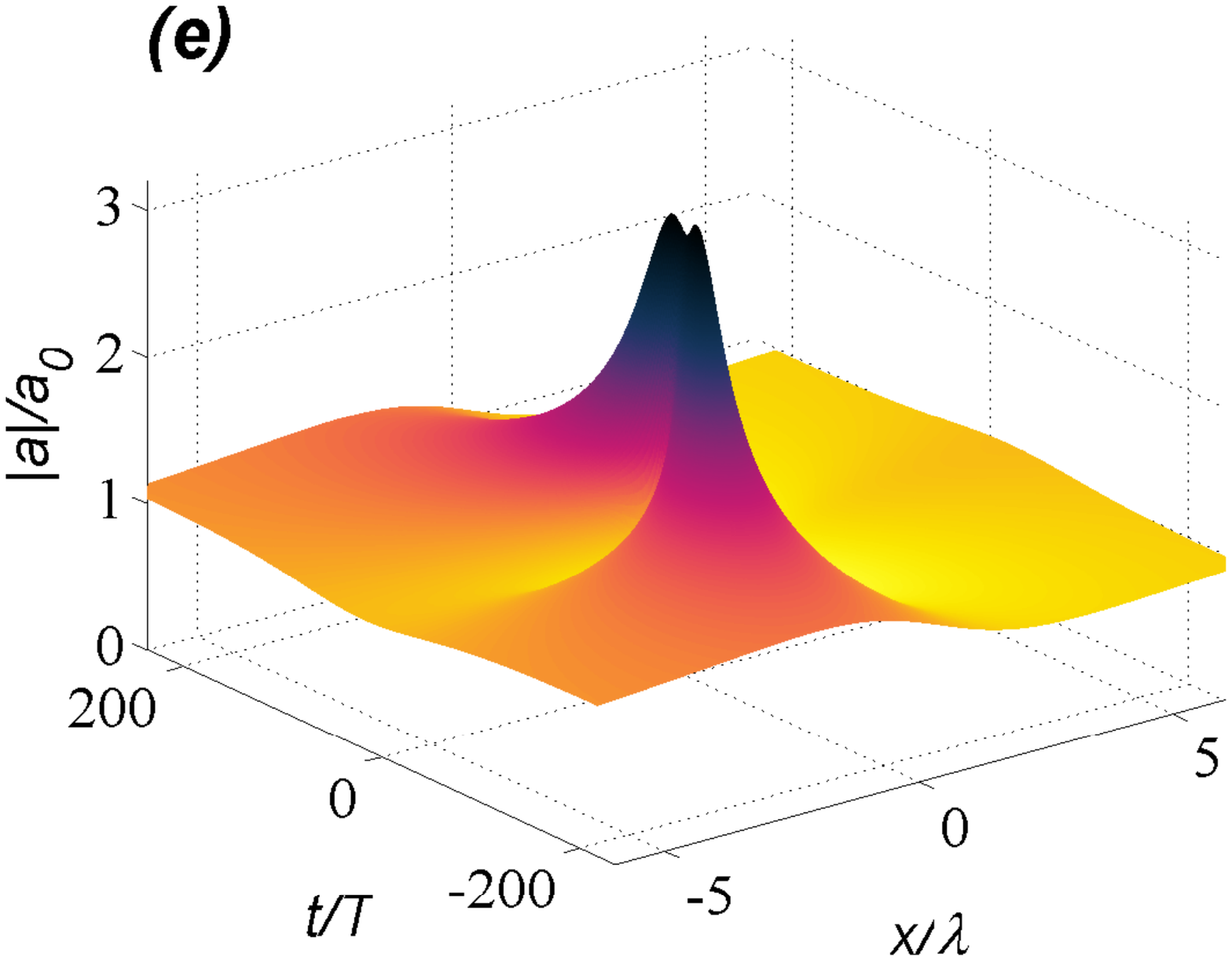}
\includegraphics[width=0.23\textwidth]{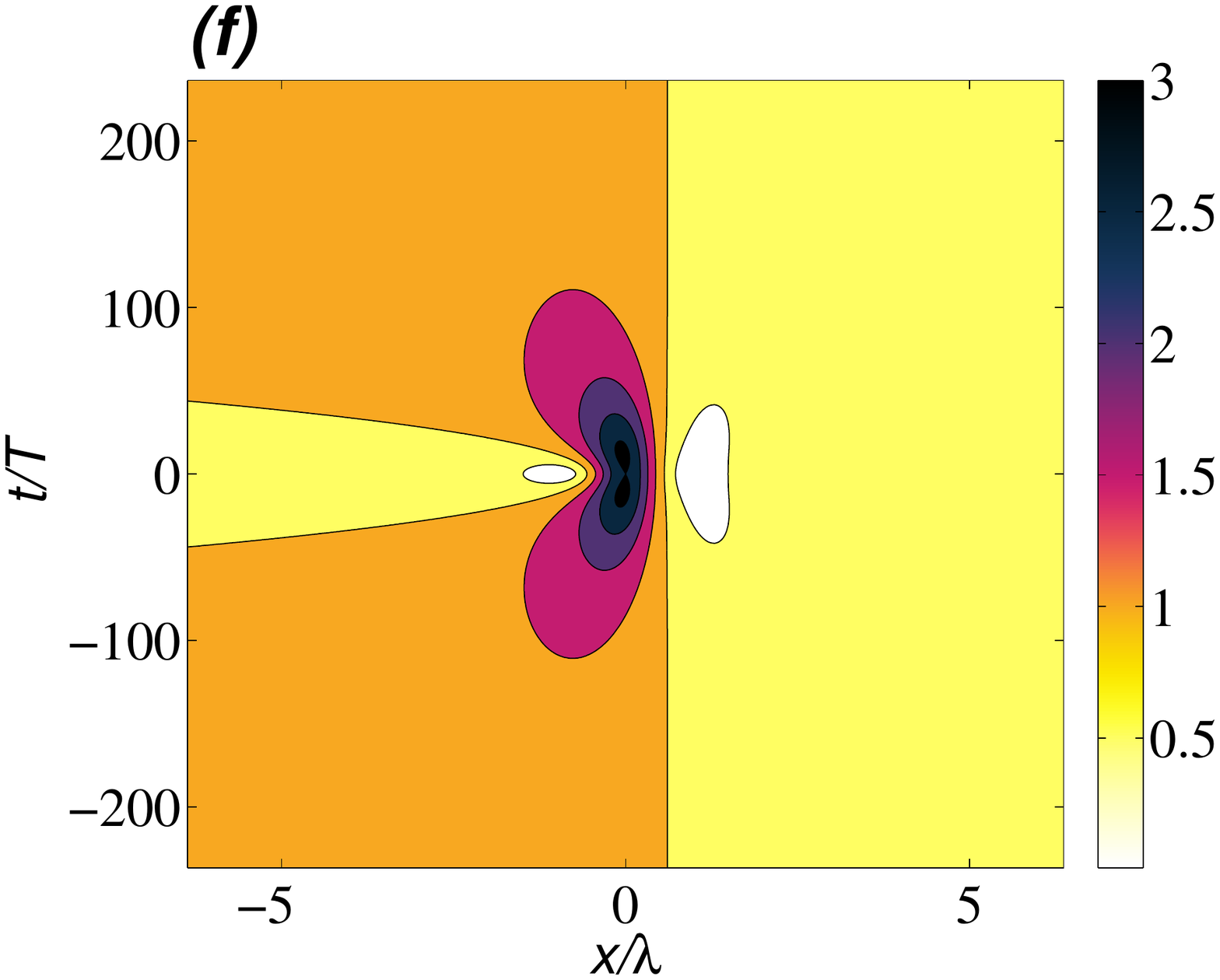}
\caption{The Peregrine solution for $a_0=1$ and $\eps = 0.1$. Panels $(a)$-$(b)$: unperturbed solution with $\Gamma_M =0$; $(c)$-$(d)$: $\Gamma_M/f = \eps$; $(e)$-$(f)$: 
$\Gamma_M/f = 2\eps$.}
\label{fig:1}
\end{figure}

The Akhmediev solution~\cite{1987TMP....72..809A}, which for large negative times corresponds to a perturbed Stokes wave, represents the nonlinear evolution of the 
modulational instability. It is periodic in space and its analytical form~\cite{2013PhR...528...47O} becomes for fast-growing waves
\ba
&&a(x,t) = a_0 \exp[-i M a_0^2 t-i\beta_3 t] \times \nonumber \\
&& \left(\frac{\sqrt{2}\tilde \nu^2 \cosh[\Omega t] - i\sqrt{2} \tilde \sigma \sinh[\Omega t]}{\sqrt{2} \cosh[\Omega t] - \sqrt{2-\tilde\nu^2} \cos[K(x-i\beta_2 t)]} -1\right)
\label{Akhmediev}
\ea    
where $\tilde \nu = K\sqrt{\beta_1/M}/a_0$ ($K$ is the wavenumber of the perturbation), $\tilde \sigma = \tilde\nu\sqrt{2-\tilde\nu^2}$ 
and $\Omega = Ma_0^2 \tilde \sigma$.
In Fig.~3 we show the unperturbed Akhmediev solution for $a_0=1$, 
$\eps = 0.1$ and $k/K = 5$ in panels $(a)$-$(b)$, 
and the wind-perturbed solution in panels $(c)$-$(d)$ and $(e)$-$(f)$ for  
$\Gamma_M/f= 0.1$ and $\Gamma_M/f= 0.17$, respectively. 
The effect of the wind is similar to the case described for the Peregrine soliton, 
as can be expected from the interrelation between the first-order solutions of the NLS equation  described in Ref.~\cite{Akhmediev2009}, although it becomes significant at a lower wind 
forcing for the same steepness. 
The shape of the wave along the direction of wave propagation is distorted,  
the maximum amplitude splits into two peaks symmetrical with respect to the $t=0$ line at 
$\Gamma_M/f = 0.136$ (for $\eps=0.1$) and then it slightly increases for larger values, 
and the lifetime increases as the growth rate increases, 
as shown in Fig.~2 (red thick-solid and dashed lines) and Fig.~3. 
The spatial periodicity is not affected. 

\begin{figure}
\centering
\includegraphics[width=0.49\textwidth]{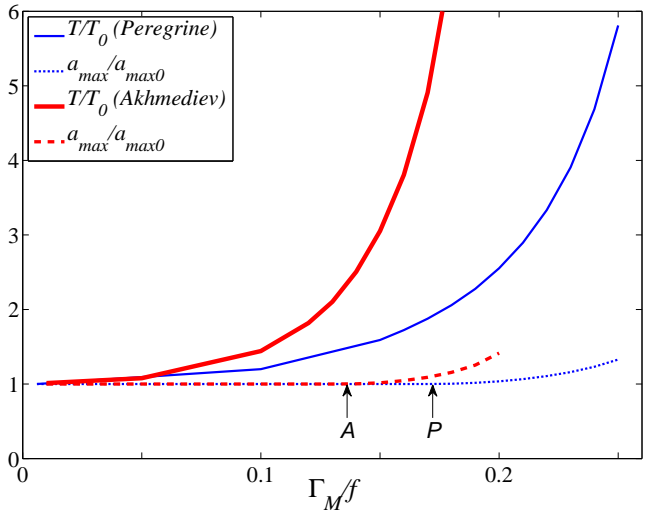}
\caption{Maximum amplitude $a_{max}/a_{max0}$ and lifetime $T/T_0$ (normalised with respect to the unperturbed values) of the Peregrine (P) soliton (blue solid and dotted lines) and the Akhmediev (A) soliton (red thick-solid and dashed lines) as a function of the growth rate $\Gamma_M/f$ for steepness $\eps = 0.1$ and $\Gamma_M/f = O(\eps)$. The position of the arrows corresponds to the value of $\Gamma_M/f$ for which the maximum splits into two peaks. }
\label{fig:2}
\end{figure}

\begin{figure}
\centering
\includegraphics[width=0.23\textwidth]{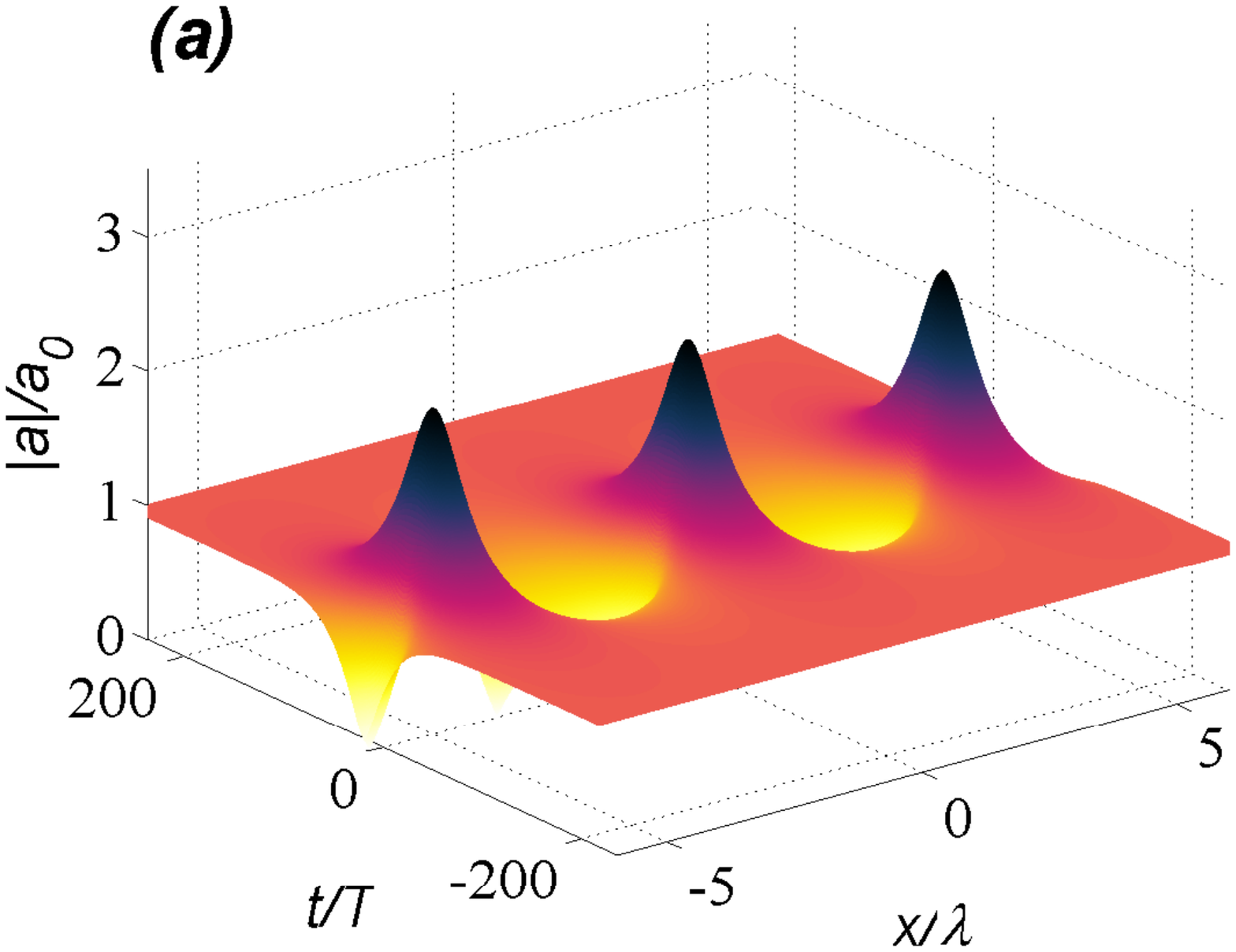}
\includegraphics[width=0.23\textwidth]{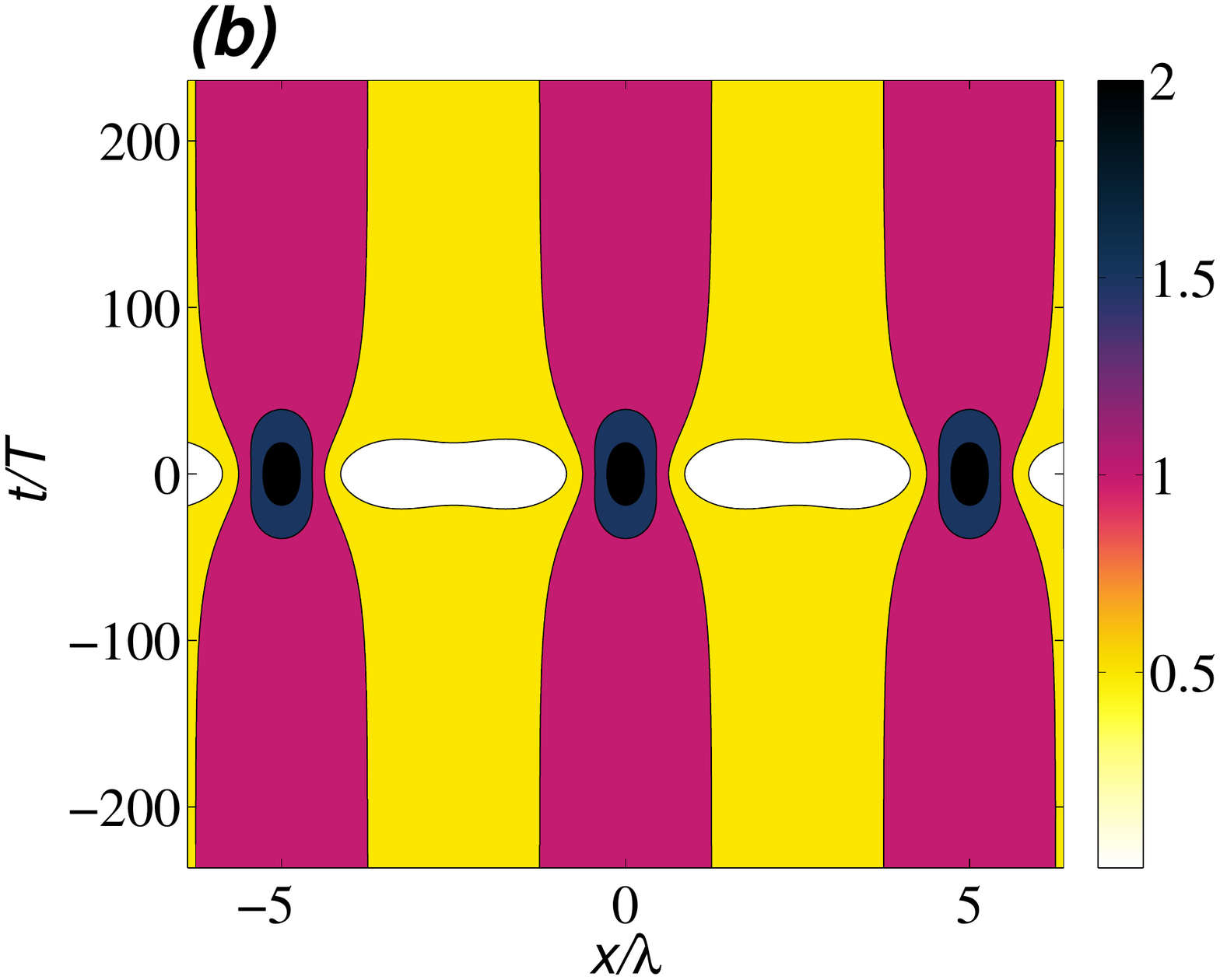}
\includegraphics[width=0.23\textwidth]{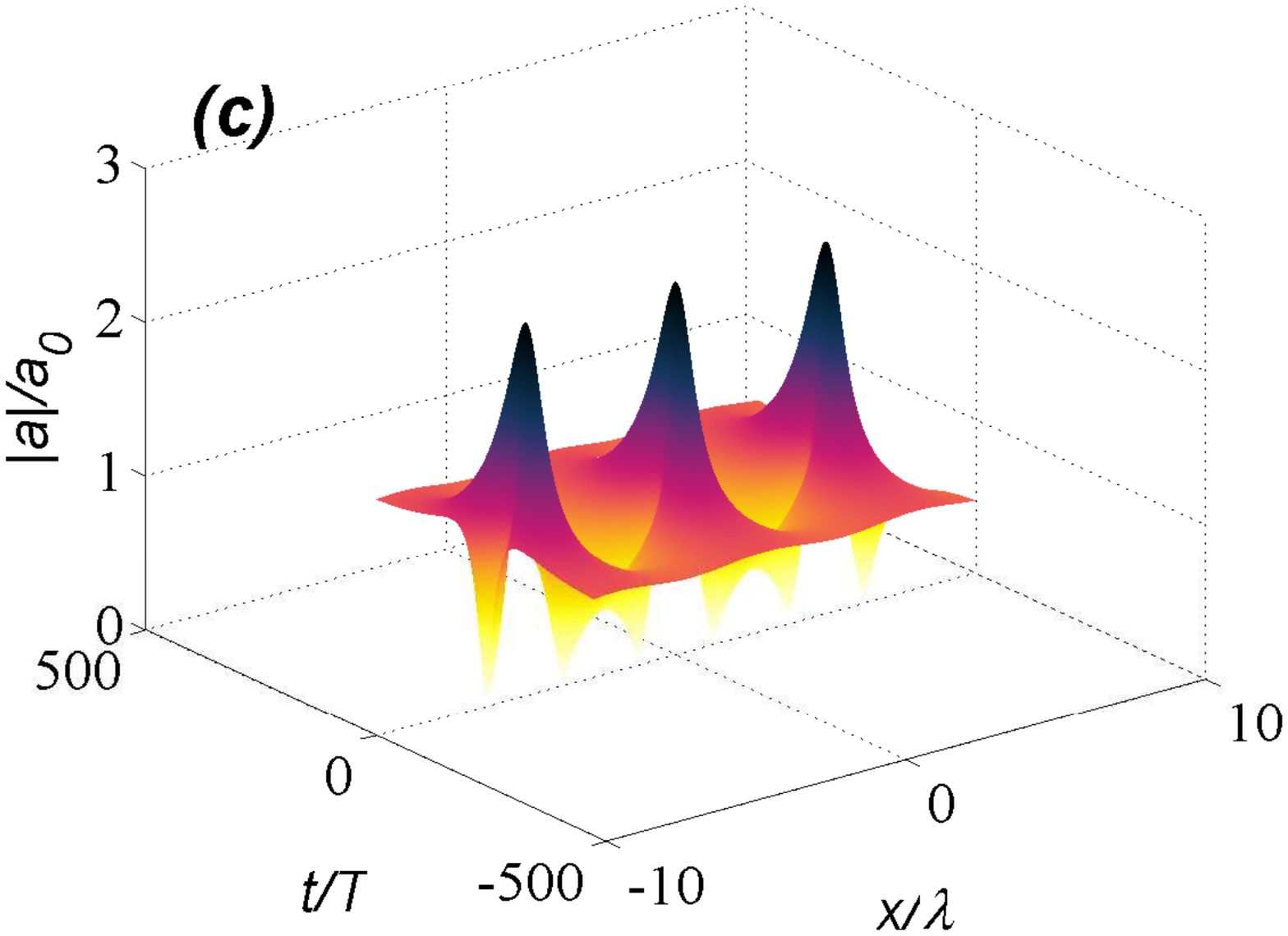}
\includegraphics[width=0.23\textwidth]{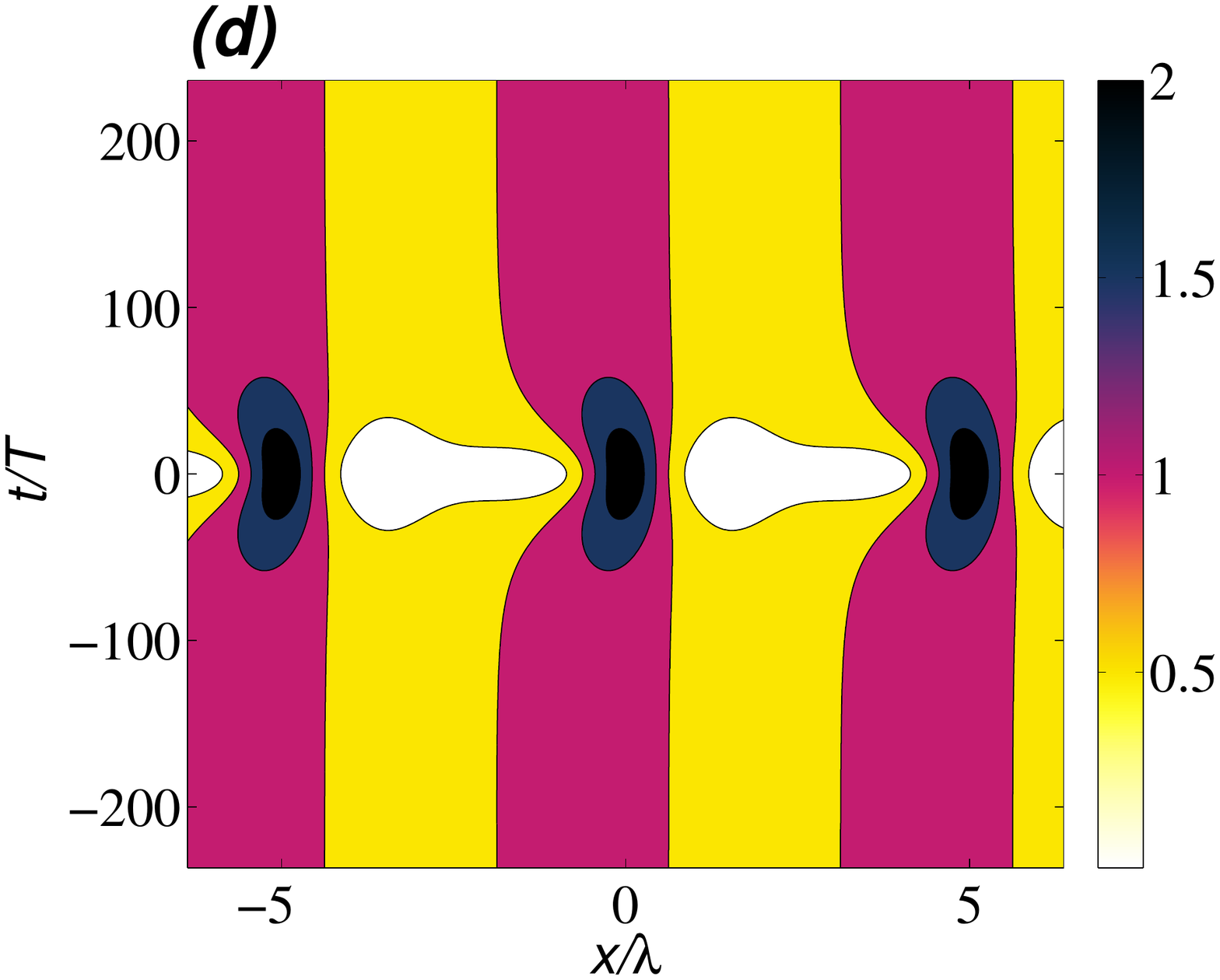}
\includegraphics[width=0.23\textwidth]{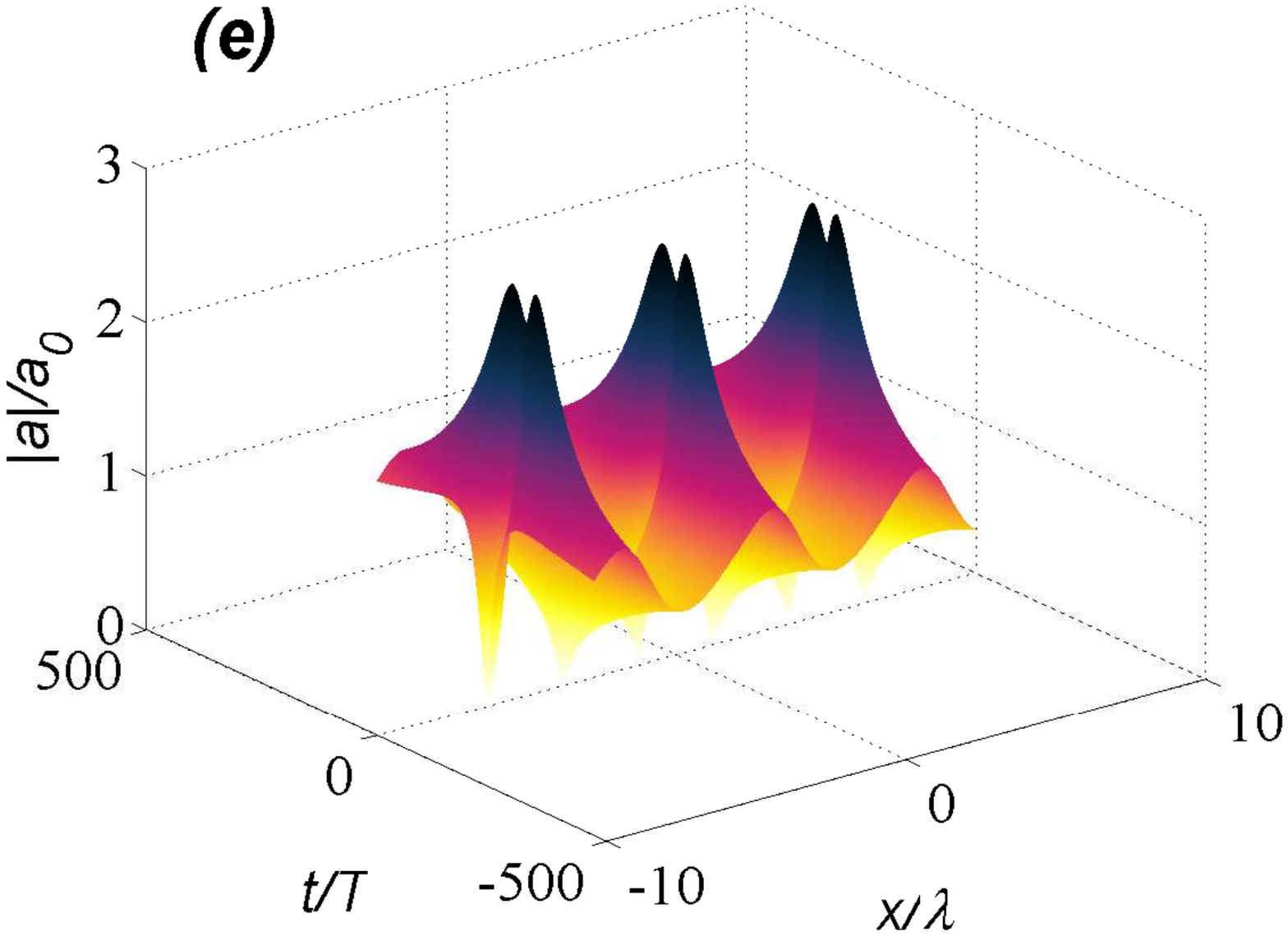}
\includegraphics[width=0.23\textwidth]{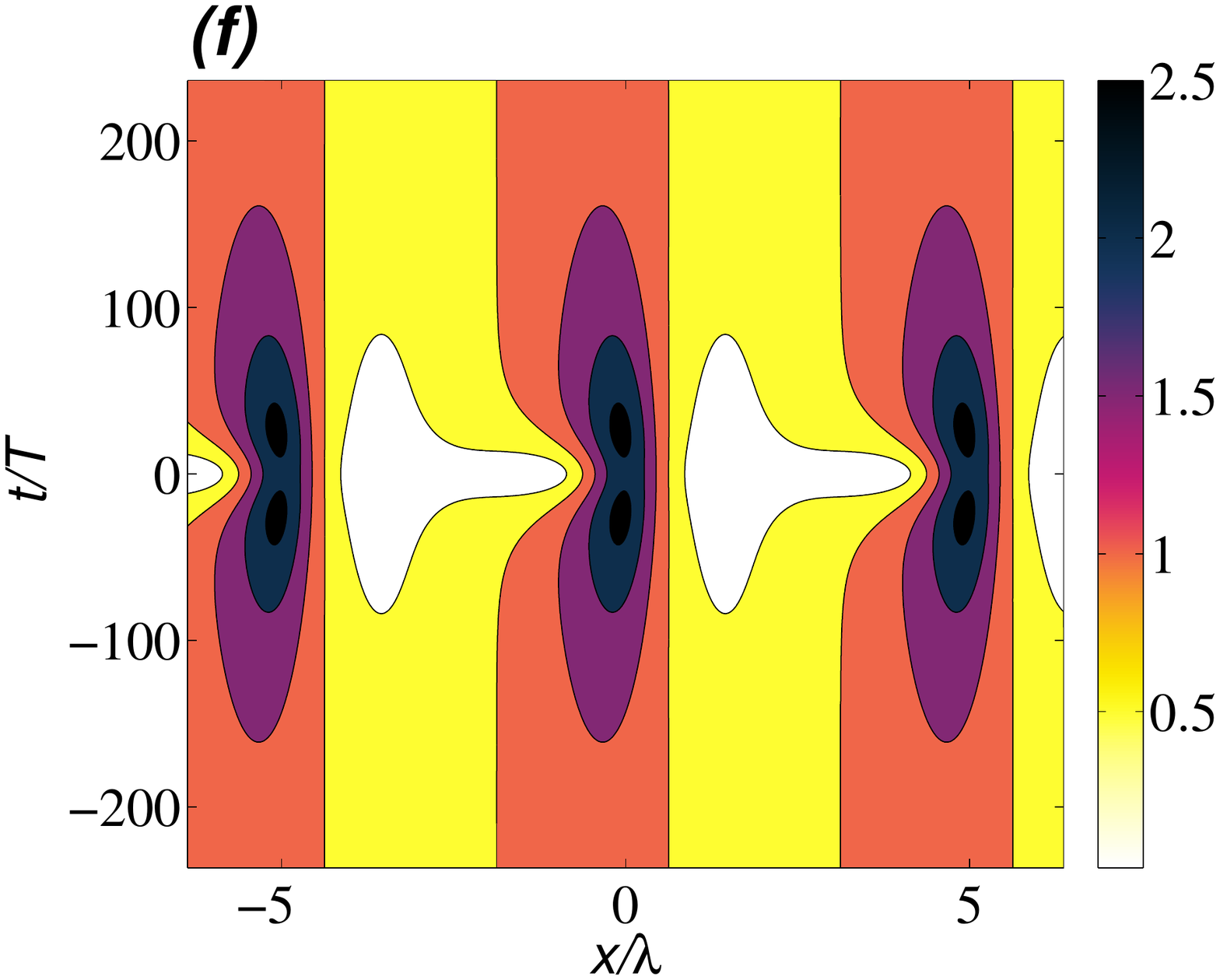}
\caption{The Akhmediev solution for $a_0=1$, $\eps = 0.1$ and $k/K = 5$. Panels $(a)$-$(b)$: 
unperturbed solution with $\Gamma_M =0$; $(c)$-$(d)$: $\Gamma_M/f =0.1$; 
$(e)$-$(f)$: $\Gamma_M/f = 0.17$.}
\label{fig:3}
\end{figure}

The Kusnetsov-Ma solution~\cite{1979StAM...60...43M} is periodic in time. While the large-time limit for the Akhmediev solution is a small perturbation of a plane wave, the Kusnetsov-Ma solution is never small and cannot grow from the modulational instability. Its analytical form~\cite{2013PhR...528...47O} becomes for fast-growing waves
\ba
&&a(x,t) = a_0 \exp[-i M a_0^2 t-i\beta_3 t] \times \nonumber \\ 
&& \left(\frac{-\sqrt{2}\tilde \mu^2 \cos[\Omega t] + i\sqrt{2} \tilde \rho \sin[\Omega t]}{\sqrt{2} \cos[\Omega t] - \sqrt{2+\tilde\mu^2} \cosh[K(x-i\beta_2 t)]} -1\right)
\label{Ma}
\ea 
where $\tilde \mu = K\sqrt{\beta_1/M}/a_0$, $\tilde \rho = \tilde \mu \sqrt{2+\tilde\mu^2}$ and 
$\Omega = Ma_0^2\tilde\rho$.
In Fig.~4 we show the unperturbed Kusnetsov-Ma soliton for $a_0=1$, $\eps = 0.1$ and 
$\tilde\mu =\sqrt{2}$ in panels $(a)$-$(b)$, and the wind-perturbed solution in panels 
$(c)$-$(d)$ for $\Gamma_M/f = 0.0135$. Even in this case, the effect of the wind is very strong: the amplitude of the wave is modulated in time with an envelope that diverges as $t\to \pm \infty$. Therefore this solution is eliminated by imposing the boundary condition that the ocean surface is unperturbed at $t= -\infty$.    

\begin{figure}
\centering
\includegraphics[width=0.23\textwidth]{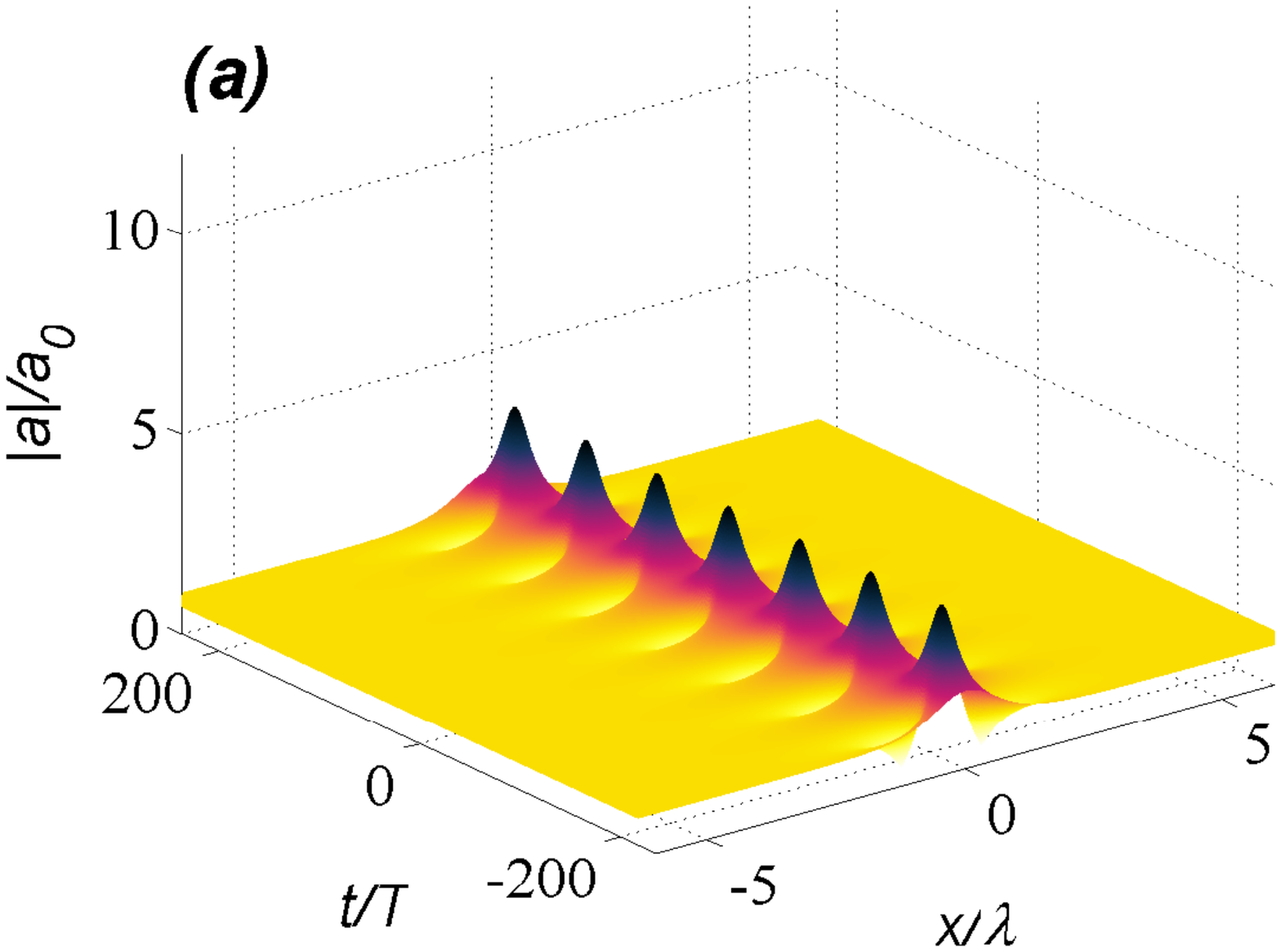}
\includegraphics[width=0.23\textwidth]{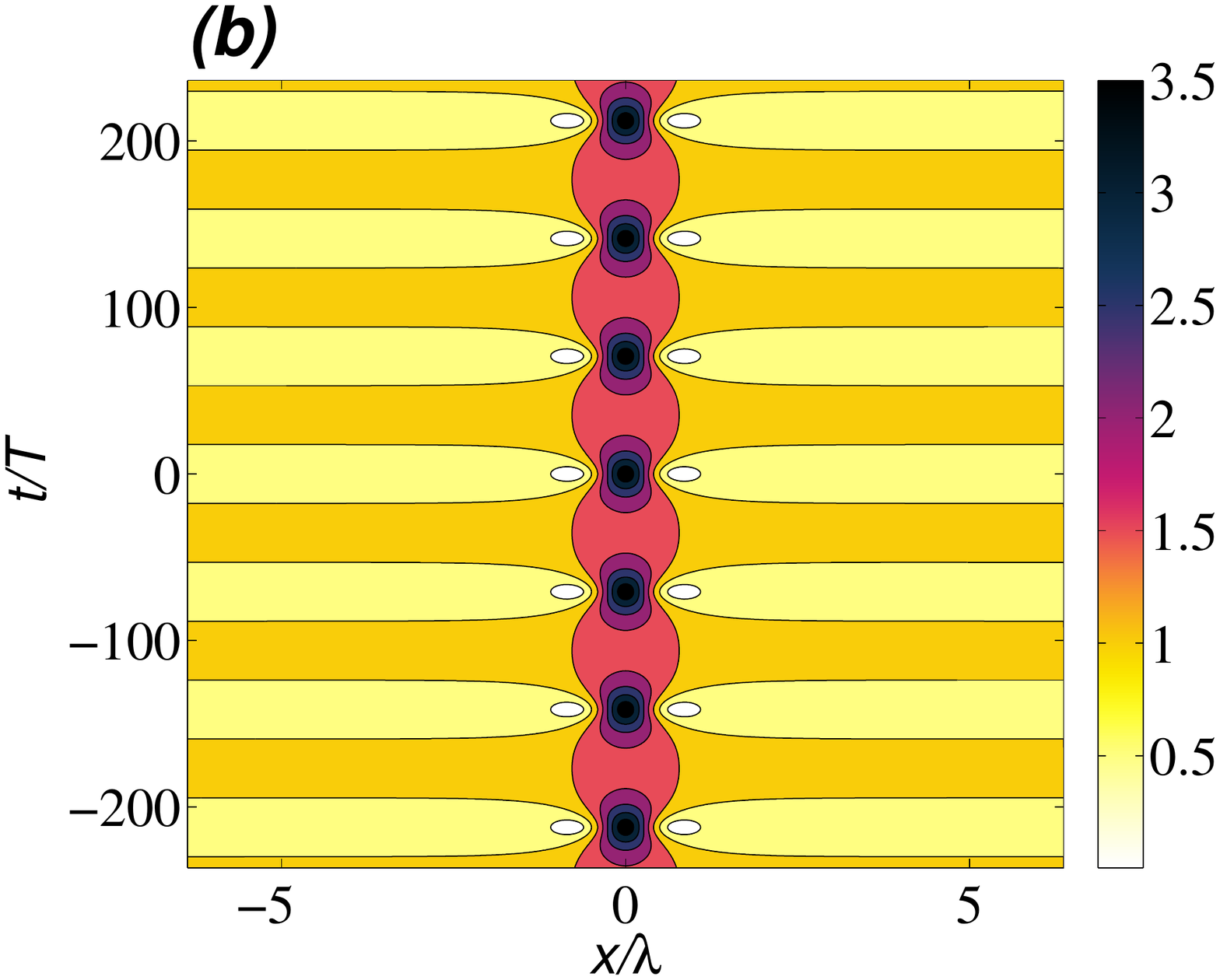}
\includegraphics[width=0.23\textwidth]{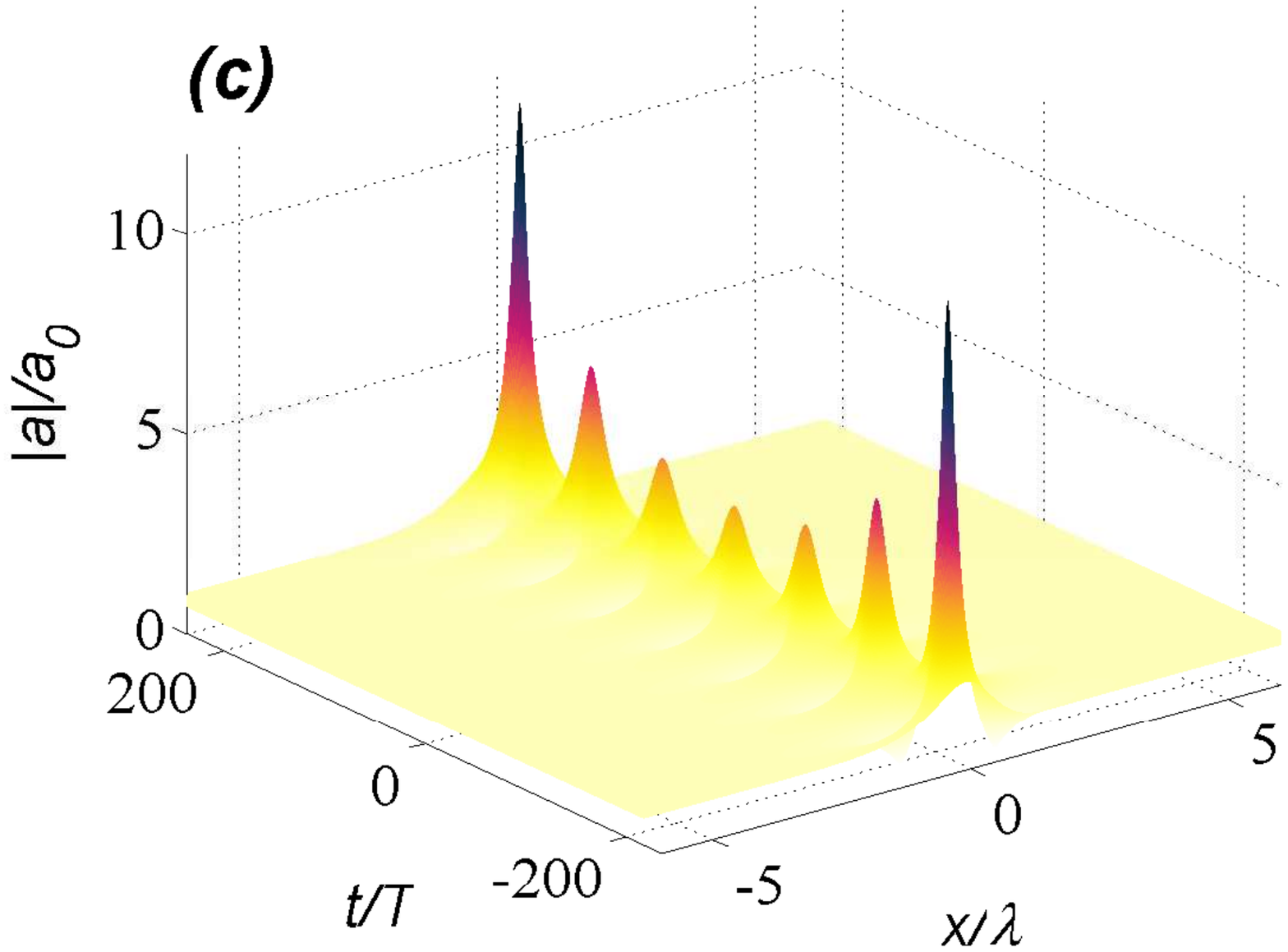}
\includegraphics[width=0.23\textwidth]{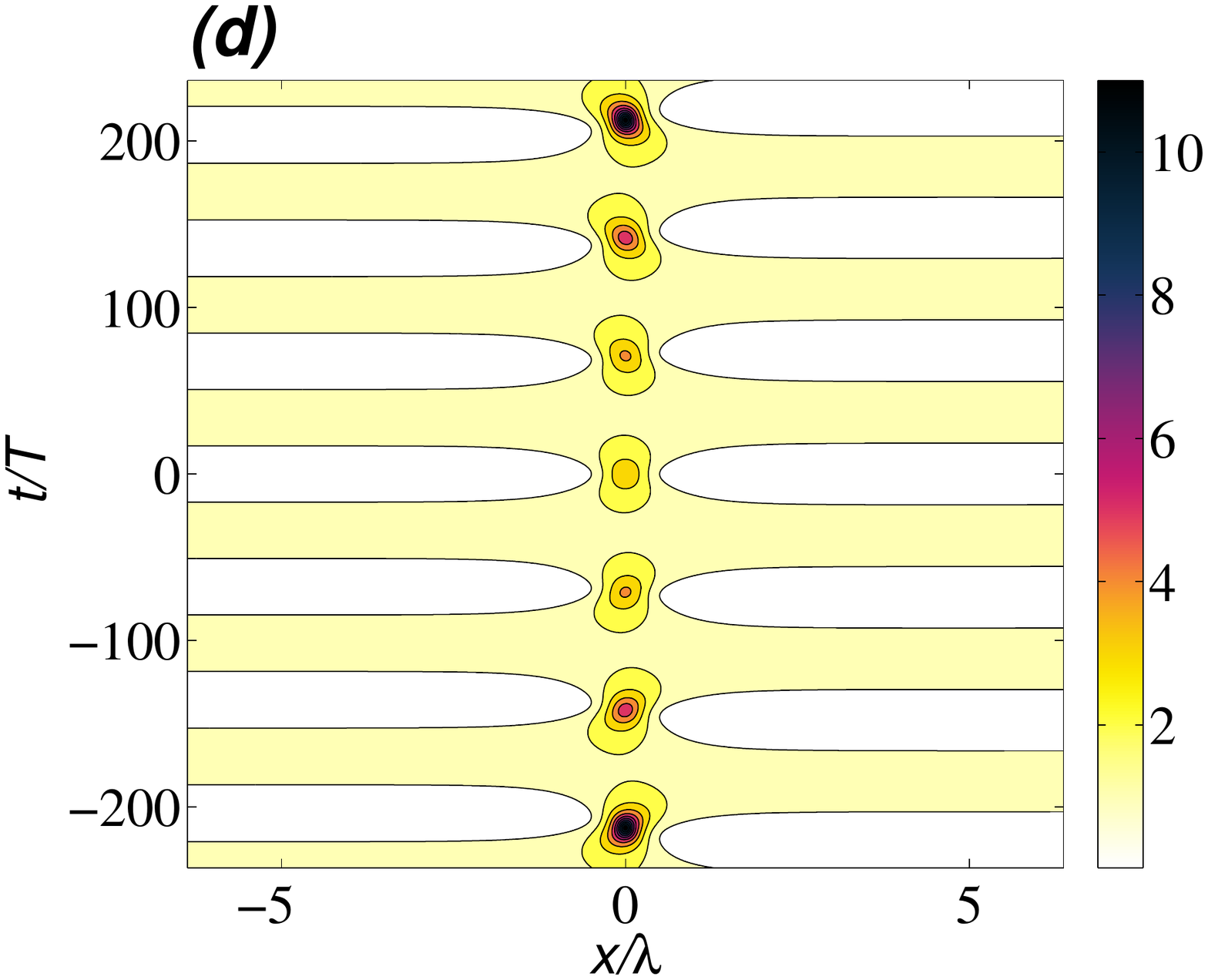}
\caption{The Kusnetsov-Ma solution for $a_0=1$, $\eps = 0.1$ and $\tilde\mu =\sqrt{2}$. 
Panels $(a)$-$(b)$: unperturbed solution with $\Gamma_M =0$; $(c)$-$(d)$: 
$\Gamma_M/f = 0.0135$.}
\label{fig:4}
\end{figure}

\section{Conclusions}
\label{concl}

The forcing term in the physical context of water surface waves is the wind. Different mechanisms have been proposed in the literature for leading to wave amplification under the action of wind. In this Letter we have considered a weakly nonlinear model (the NLS equation)  and the Miles mechanism for the wind-wave coupling. The growth rate $\Gamma_M/f$ of the wave energy has been compiled by different authors~\cite{Banner2002,Farrell2008} and its values range from $10^{-3}$-$10^{-2}$ for fast-moving waves ($c_p/u^*> 5$) to $10^{-2}$-1 for slow-moving waves and laboratory tank experiments ($c_p/u^*\le 5$).
This prompted us to investigate the case $\Gamma_M/f = O(\eps)$, where $\eps$ is the wave steepness, which both in ocean and in tank experiments is of the order of 0.1 or less. 

We have used the method of multiple scales for deriving the wind-forced NLS equation with wave growth rate of first order in the wave steepness. Beside wave amplification, the effect of the wind is to modify the dispersion term. A simple coordinate transformation reduces the wind-forced NLS 
equation into the standard NLS equation with constant coefficients. 
We have thus shown that the 
soliton solutions (Peregrine, Akhmediev and Kuznetsov-Ma solutions) are modified in the presence of wind. In particular, the lifetime of both the Peregrine and the Akhmediev solitons increases for large growth rates. We find that the maximum amplitude of these solitons slightly increases for growth rates larger than a certain value, characterised by the fact that two maxima appear at opposite positions with respect to the $t=0$ line. The enhancement of both lifetime and maximum amplitude of rogue waves under the action of wind has been observed in  tank experiments and numerical simulations of dispersive focusing~\cite{2006EJMF...25..662T,2008NPGeo..15.1023T}, thus confirming the relevance in this context of the case $\Gamma_M/f=O(\eps)$ with respect to $\Gamma_M/f=O(\eps^2)$ for which the soliton lifetime does not change under the action of wind~\cite{OnoratoProment2012}.

The results presented here should be tested in wind-wave tank experiments with different ranges of growth rate and steepness to characterise the transition between the two different regimes 
$\Gamma_M/f=O(\eps)$ and $\Gamma_M/f=O(\eps^2)$. 

\medskip
\noindent
{\small We would like to thank Jean-Pierre Wolf and Martin Beniston for interesting discussions and the two anonymous referees for useful comments. We acknowledge financial support from the ERC advanced grant "Filatmo" and the CADMOS project.}



\bibliographystyle{elsarticle-num}
\bibliography{waves.bib}


\end{document}